\documentclass[8.5pt,twoside,twocolumn]{article}

\usepackage[super,sort&compress,comma]{natbib} 
\usepackage{times,mathptmx}
\usepackage{sectsty}
\usepackage{balance} 
\usepackage{bm}
\usepackage{graphicx} 
\usepackage{lastpage}
\usepackage[format=plain,justification=raggedright,singlelinecheck=false,font=small,labelfont=bf,labelsep=space]{caption} 
\usepackage{fancyhdr}
\usepackage{xcolor}
\newcommand{\vt}[1]{\mathbf{#1}}

\begin{document}

\makeatletter 
\def\subsubsection{\@startsection{subsubsection}{3}{10pt}{-1.25ex plus -1ex minus -.1ex}{0ex plus 0ex}{\normalsize\bf}} 
\def\paragraph{\@startsection{paragraph}{4}{10pt}{-1.25ex plus -1ex minus -.1ex}{0ex plus 0ex}{\normalsize\textit}} 
\renewcommand\@biblabel[1]{#1}            
\renewcommand\@makefntext[1]%
{\noindent\makebox[0pt][r]{\@thefnmark\,}#1}
\makeatother 
\renewcommand{\figurename}{\small{Fig.}~}
\sectionfont{\large}
\subsectionfont{\normalsize} 


\twocolumn[
  \begin{@twocolumnfalse}
\noindent\LARGE{\textbf{Buckling dynamics of a solvent-stimulated stretched elastomeric sheet$^\dag$}}
\vspace{0.6cm}

\noindent\large{\textbf{Alessandro Lucantonio,$^{\ast}$\textit{$^{a}$} Matthieu Roch\'{e},\textit{$^{b\ddag}$} Paola Nardinocchi\textit{$^{a}$}, and
Howard A.~Stone\textit{$^{b}$}}}\vspace{0.5cm}

\noindent\textit{\small{\textbf{Received Xth XXXXXXXXXX 20XX, Accepted Xth XXXXXXXXX 20XX\newline
First published on the web Xth XXXXXXXXXX 200X}}}

\noindent \textbf{\small{DOI: 10.1039/b000000x}}
\vspace{0.6cm}

\noindent \normalsize{When stretched uniaxially, a thin elastic sheet  may exhibit buckling. The occurrence of buckling depends on the geometrical properties of the sheet and the magnitude of the applied strain. Here we show that an elastomeric sheet initially stable under uniaxial stretching can destabilize when exposed to a solvent that swells the elastomer. We demonstrate experimentally and computationally that the features of the buckling pattern depend on the magnitude of stretching, and this observation offers a new way for controlling the shape of a  swollen homogeneous thin sheet.}
\vspace{0.5cm}
 \end{@twocolumnfalse}
  ]

\footnotetext{\dag~Electronic Supplementary Information (ESI) available: description of the method used for reconstructing the shape of the swollen sheet and additional experimental data on the dynamics of wrinkling. See DOI: 10.1039/b000000x/}


\footnotetext{\textit{$^{a}$~Dipartimento di Ingegneria Strutturale e Geotecnica, Sapienza Universit\`{a} di Roma, Rome, Italy. E-mail: alessandro.lucantonio@uniroma1.it}}
\footnotetext{\textit{$^{b}$~Department of Mechanical and Aerospace Engineering, Princeton University, Princeton NJ, USA.}}


\footnotetext{\ddag~Now at: Laboratoire de Physique des Solides, Universit\'{e} Paris Sud - CNRS UMR 8502, B\^{a}timent 510, 91405 Orsay, France.}

\section*{Introduction}

Soft materials such as tissues and gels can become mechanically unstable when experiencing a volumetric expansion such as swelling or growth \cite{Tanaka1987,AmarGoriely2005}. These instabilities constrain the shape of both natural and engineered structures, such as the gut \cite{Savin2011}, bacterial biofilms \cite{Trejo2013}, multilayered materials \cite{Mora2006,Sultan2008} and composite media \cite{Wu2013}. The study of swelling-induced instabilities in heterogeneous thin sheets made of gels and elastomers with spatially varying chemical composition has led to new design principles to control the shape of soft matter systems \cite{Shu2010,Kim2012,Wu2013}.

Here, we show that shape control can be achieved in chemically homogeneous elastomeric sheets experiencing swelling by applying a pre-strain. We document experiments showing that a thin free-standing sheet, stretched between clamps and mechanically stable, destabilizes by forming wrinkles when swollen with a good solvent for the elastomer. In contrast to the spontaneous wrinkling of uniaxially stretched elastomeric sheets \cite{Friedl2000,Cerda2002,Puntel2011}, the instability is triggered here by a non-mechanical stimulus that is swelling due to the migration of solvent within the elastomer. The wavelength and the amplitude of the buckling pattern are a function of the initial strain applied to the sheet. We follow the growth of the buckling pattern with time and we show that its wavelength is selected at the moment the sheet becomes unstable while the amplitudes of the wrinkles increases with time. We rationalize the properties of the dynamic buckling pattern using a numerical stress-diffusion model that relates the stress state of the elastic sheet to the amount of solvent absorbed by the elastomeric matrix. The numerical model is in good agreement with the experimental data, thus supporting the connection we propose between the mechanics of the instability and the thermodynamics of swelling. 

\section*{Experimental}

We performed experiments with silicone sheets (Marian Chicago Inc.) with length $L_0=50\ $mm, width $W_0=60\ $mm, and thickness $h_0=250\ \mu$m at rest. The sheets had Young's moduli $E=1.4$ MPa (Shore 35A), $E=0.73$ MPa (Shore 20A), $E=0.36$ MPa (Shore 10A) \cite{Meier2011}, and both of their ends were plasma-bonded to glass slides before being sandwiched using other glass slides glued with epoxy resin. We checked that the sheets remained bonded to the glass slides for the whole duration of all of the experiments we performed.

\begin{figure*}[!htbp]
\begin{center}
\includegraphics[scale=0.9]{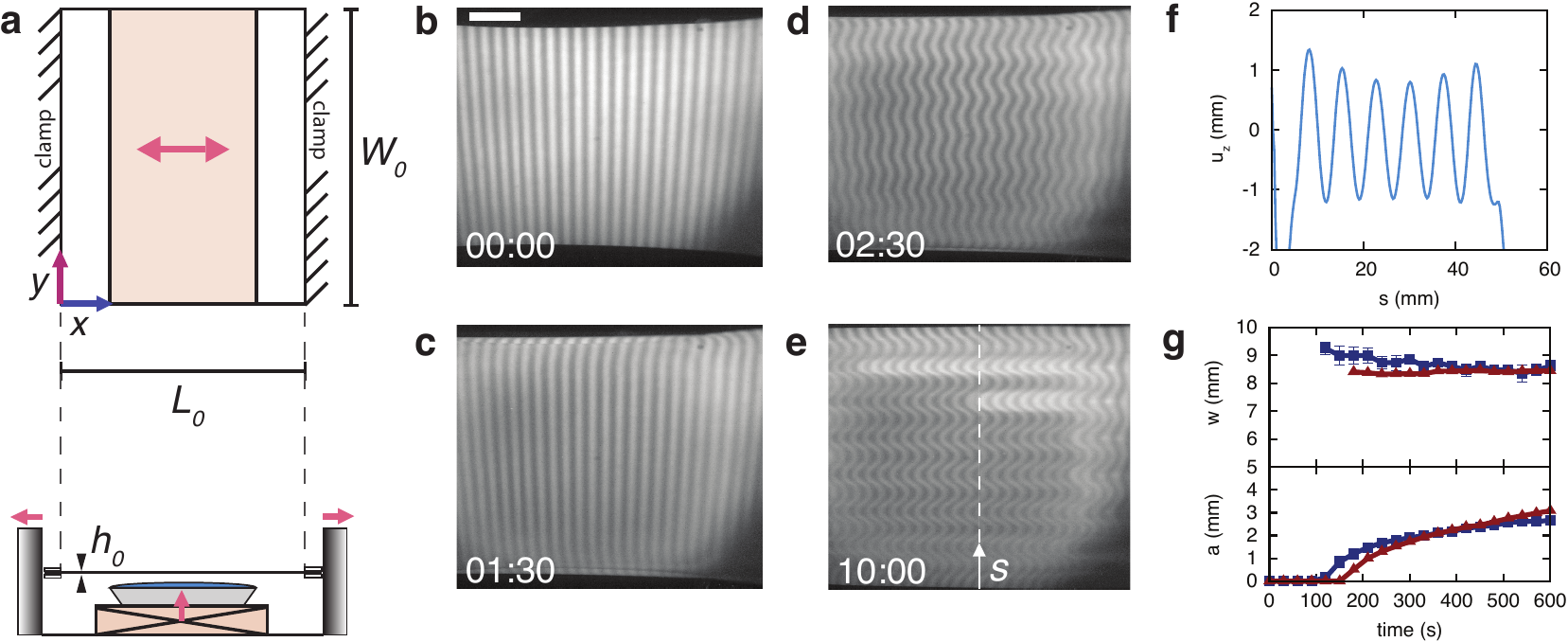}
\caption{(a) Geometry of the experimental setup. The shaded area corresponds to the region of the bottom surface of the sheet where the solvent flux is prescribed in the numerical simulations. The horizontal arrows at the center of the sheet indicate the direction of stretching. (b)-(e) Frames showing the state of the sheet at different times during swelling. Initial strain $\varepsilon = 0.9$, scale bar: 10 mm. (see also Movie S1$^\dag$). (f) Raw data of the vertical displacement $u_z$ reconstructed via WTP along the dashed line in (e). (g) Dynamics of the wavelength $w$ (top) and amplitude $a$ (bottom) of the wrinkles for $\varepsilon = 0.45$; squares:  experimental average values along the width at $x=L_0/2$, triangles: numerical results.}
\label{fig:sheet_geometry}
\end{center}
\end{figure*}

Clamps that could move along the longitudinal axis of the sheet held the glass slides and  allowed us to vary the stretched length $L$ of the sheet beyond its length $L_0$ at rest. Hence, we defined the nominal longitudinal strain $\varepsilon = (L-L_0)/L_0$ in the range \mbox{$0.3\leq\varepsilon\leq 0.9$}, over which all the sheets remained mechanically stable, \textit{i.e.} no wrinkles appeared. A typical experiment started when the sheet was placed in contact with the surface of a square bath (side $\ell=43$ mm) of silicone oil (Sigma Aldrich, kinematic viscosity $\eta=2$ cSt, density \mbox{$\rho = 873$ kg/m$^{ 3}$}), which is a good solvent for the elastomer (Fig.~\ref{fig:sheet_geometry}a). The sheets ruptured for $\varepsilon>1$ during swelling.

We reconstructed the dynamic shape of the swollen sheet using wavelet transform profilometry (WTP) \cite{Zhong2004}. We used a photographic lens (fixed focal length \mbox{$f = 28$ mm}) to project a fringe pattern (vertical stripes in \mbox{Fig.~\ref{fig:sheet_geometry}b--e}) with sinusoidal intensity on the top surface of the sheet. A digital camera making an angle of $\pi/4$ with the plane of the sheet captured images of the fringe patterns every \mbox{30 s} during swelling. We processed images with a custom MATLAB code that performed the wavelet transform of a line of pixels selected by the user and common to all images (see the ESI$^\dag$ for details on the method). We were able to resolve out-of-plane deformations of the sheet with a precision of $50\ \mu$m across the entire width of the sheet, which is on the order of 45 mm depending on the initial strain. An example of a reconstructed profile using WTP is given in Fig.~\ref{fig:sheet_geometry}f. We weighed the sample before and after solvent exposure to measure the solvent mass uptake with a precision of \mbox{$1$ mg}. 

\section*{Results and discussion}

Shortly after the sheet was brought into contact with the surface of the oil bath, we observed that the free edges of the sheet expanded outwards and then curled (Fig.~\ref{fig:sheet_geometry}c, Movie S1$^\dag$). After a few hundred seconds, a time-evolving buckling pattern developed in the direction transverse to stretching (Fig.~\ref{fig:sheet_geometry}d). The system selected a wavelength \textit{w} for the wrinkles at the time at which buckling appeared. This wavelength remained nearly constant until the end of the experiment while the amplitude \textit{a} of the wrinkles continuously grew (Fig.~\ref{fig:sheet_geometry}g). The small decrease in wavelength ($< 10\%$) in time comes from the growth of new wrinkles at the edges of the sheet that compress the other wrinkles toward the center (see Movie S1$^\dag$). These transient trends for the amplitude and wavelength of the wrinkles are common to all the experiments we performed (see the ESI$^\dag$). We ran each experiment for $t_f = 600$ s; for longer times, the oil dewetted from the bottom surface of the sheet, and the exposure of the sheet to the solvent changed. We note that we were unable to identify a buckling pattern for $\varepsilon<0.3$. Also, we were not always able to track the deformation of the edges: the expansion of the sheet moved the edges to locations where the reference fringe pattern was not defined, preventing us from reconstructing the sheet profile.

\begin{figure*}[!t]
\begin{center}
\includegraphics[scale=1]{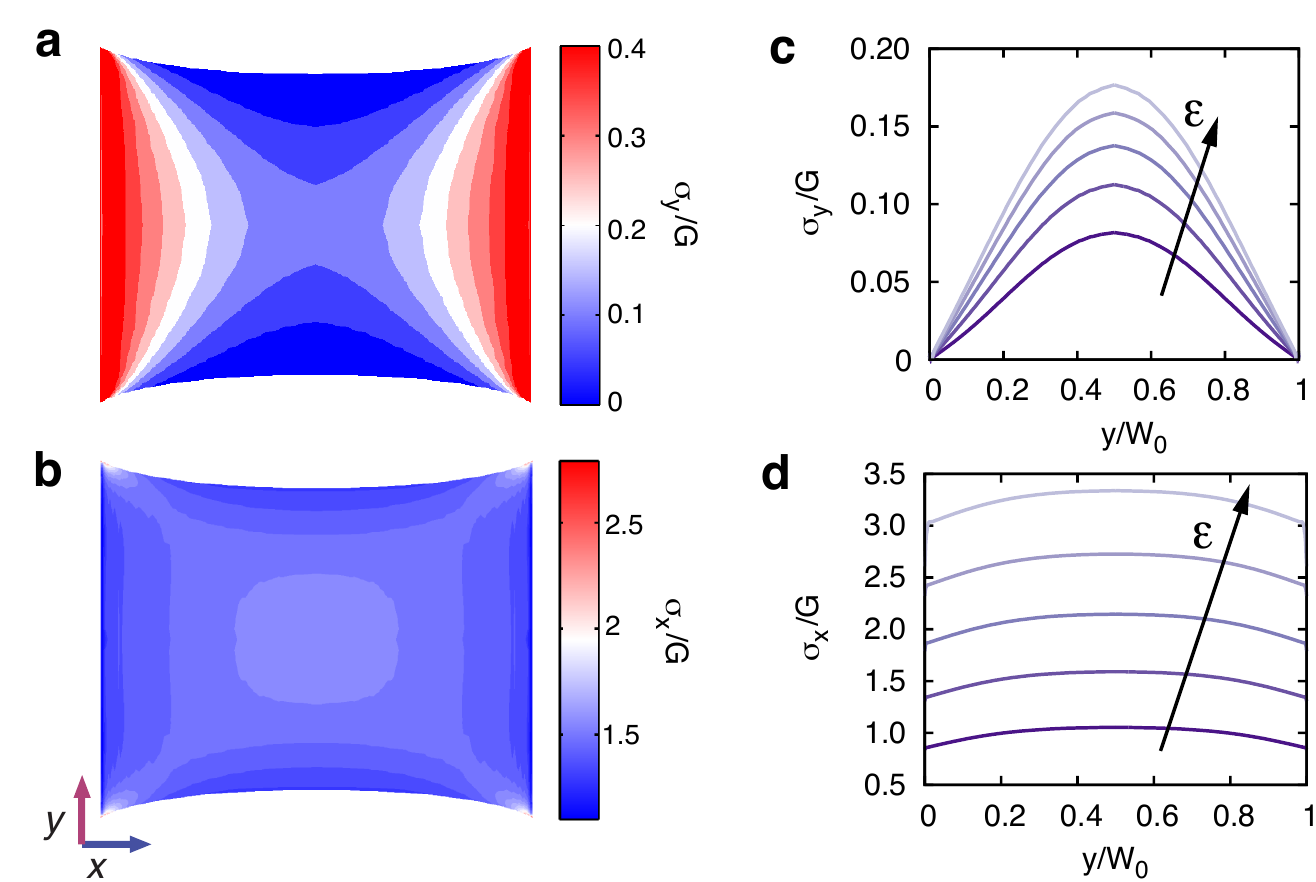}
\caption{Stress fields extracted from numerical simulations for stretching of a dry sheet. (a) Contour plots of the transverse non-dimensional stress $\sigma_y/G$ and (b) the longitudinal non-dimensional stress $\sigma_x/G$ on the midplane of the sheet for \mbox{$\varepsilon = 0.45$}. (c) Transverse non-dimensional stress $\sigma_y/G$ and (d) longitudinal non-dimensional stress $\sigma_x/G$ along a segment parallel to the $y$ axis located at $x=L_0/2$, $z=h_0/2$ (midplane). The lines correspond to different applied strains $\varepsilon$: 0.3, 0.45, 0.6, 0.75, 0.9.}
\label{fig:contours_stress_preload}
\end{center}
\end{figure*}

To gather insight on the mechanisms that lead to the formation of wrinkles and drive the evolution of the pattern, we performed a set of finite element analyses based on a non-linear three-dimensional model for swelling gels developed and used by some of us in previous studies of swelling-induced deformations of elastomers \cite{AL2013}. Here, we focus on the simulation of the experiment performed with a sheet of Shore 35A elastomer at $\varepsilon = 0.45$, which is representative of the behaviors we observed for the other strains (see the ESI$^\dag$). In the model, we assume that the elastomer is incompressible; thus, we take the shear modulus to be \mbox{$G = E/3 \simeq 467$ kPa}. As experiments were carried out at room temperature, we set \mbox{$T = 298$ K}. The model also requires a value for the solvent molar volume $\Omega$. The silicone oil we used had a molecular weight \mbox{$M = 0.41\times10^{-3}$ kg/mol}, thus giving \mbox{$\Omega= M/\rho = 4.7\times10^{-4}$ m$^3$/mol}. Finally, we also need a dimensionless measure of the enthalpy of polymer-solvent mixing which we set to $\chi=0.2$, a typical value for a polymer in a good solvent \cite{Mark1999a}. The diffusivity of the solvent $D$ was used as a calibration parameter to fit the time-scale of swelling for $\varepsilon = 0.3$ and then held fixed for the other strains. We found that a value of \mbox{$D=0.8\times10^{-9}$ m$^2$/s} led to the best fit to the experimental data. This value of $D$ compares well with the value $D_{th}$ predicted based on the properties of the elastomer and the oil, \mbox{$D_{th}\simeq Ek/\eta\simeq 10^{-9}$ m$^2$/s}, with \mbox{$k\simeq 10^{-18}$ m$^2$} a typical value of the permeability of the elastomeric matrix \cite{Mark1999a}.

First, we characterize numerically the state of the stretched sheet in the absence of solvent. The results of the simulation for $\varepsilon=0.45$ show that transverse stresses are tensile everywhere in the sheet (Fig.~\ref{fig:contours_stress_preload}a), which prevents the growth of wrinkles, in excellent agreement with our experimental observations and past reports \cite{Zheng2009,Nayyar2011} for sheets with initial aspect ratios $L_0/W_0 <1$. The stress patterns for the other applied strains in the range $0.3 \leq \varepsilon \leq 0.9$ are qualitatively similar. In particular, both the longitudinal and the transverse stress along the width increase with the applied strain (Fig.~\ref{fig:contours_stress_preload}c--d). 


Next, we take the mechanical state of the stretched sheet in the absence of solvent as the initial configuration of our transient analysis of the solvent-induced deformations. We define the solvent flux entering the elastomer as $Q_s=m_s/S_w\,t_f$, where $m_s$ is the total solvent mass uptake at the end of one experiment, $S_w=\ell\,W$ the area of the wetted surface, and $t_f$ is the duration of a single experiment. We assume that $Q_s$ is homogeneous over the area of the wetted region of the bottom face of the sheet. The early curling of the free edges suggests that, when the dry sheet is wider than the dish (in the experiments, for $\varepsilon \leq 0.6$), the solvent rapidly spreads along the width of the sheet to reach the edges. Hence we assume that the wetted region extends all along the width of the sheet.  We were then able to capture both qualitatively and quantitatively the features of the experiments.

\begin{figure*}[!t]
\begin{center}
\includegraphics[scale=0.9]{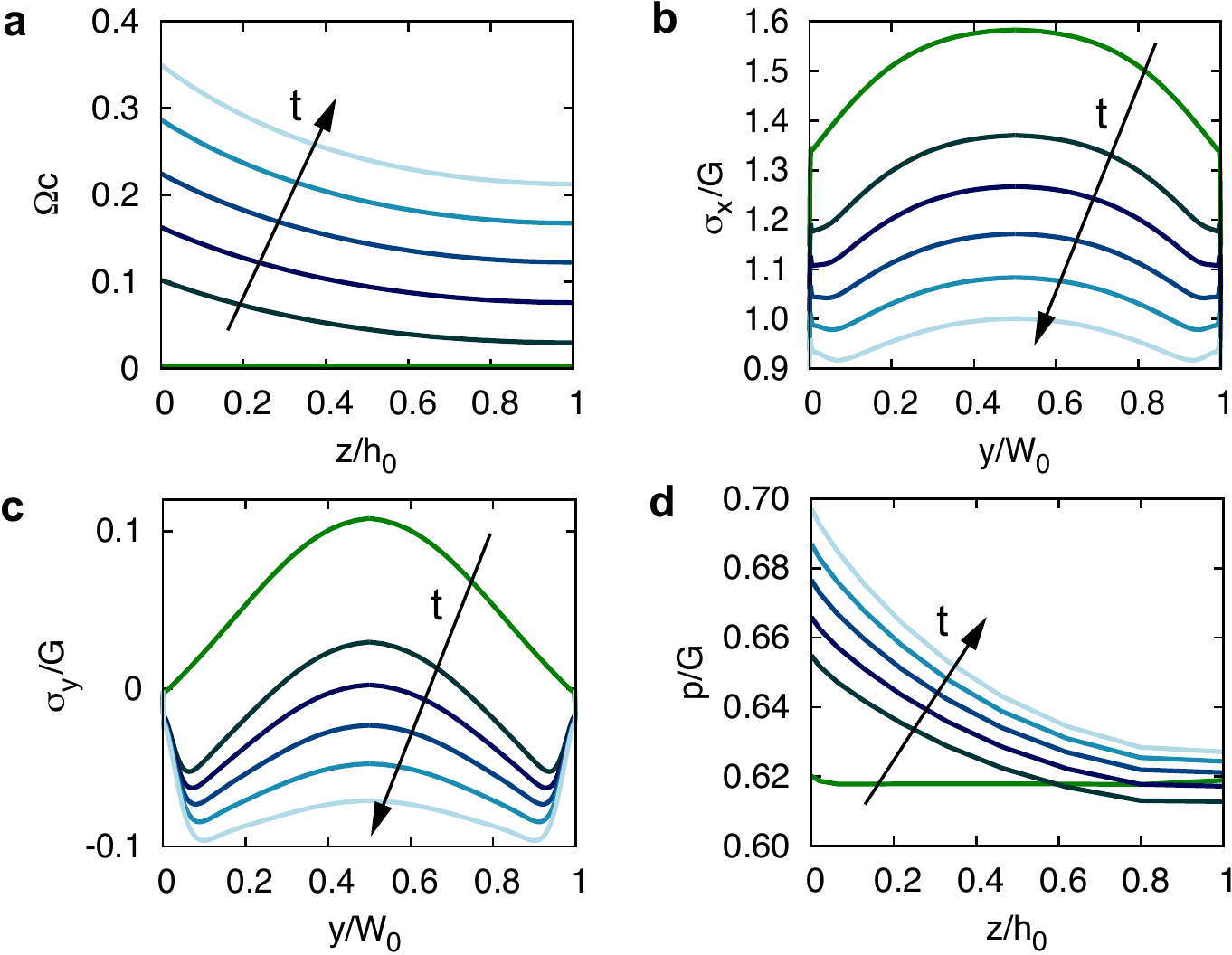} 
\caption{Mechanical state of the sheet in numerical simulations. (a) Solvent volume fraction  along the thickness $z$ at the center of the elastomeric sheet for $\varepsilon = 0.45$  ($z/h_0=0$ is the surface in contact with the solvent). The lines correspond to different times (0 to 100 s with 20 s increments) during the transient swelling process. Longitudinal stress $\sigma_x$ (b) and transverse stress  $\sigma_y$ (c) along a segment parallel to the $y$ axis located at $x/L_0 = 0.5, z/h_0=0$, for $\varepsilon = 0.45$. (d) Pressure distribution along the thickness at the center of the elastomeric sheet for $\varepsilon = 0.45$.}
\label{fig:p_along_thick}
\end{center}
\end{figure*}

Our numerical results show that the solvent concentration field $c$ right after the beginning of solvent absorption is non-homogeneous through the thickness of the sheet, with the concentration being higher in the layers closer to the solvent bath, as shown in  Fig.~\ref{fig:p_along_thick}a for the thickness at the center of the sheet. We have checked that the concentration profile along the thickness of the sheet remains qualitatively the same at other locations far from the clamps. The non-homogeneous concentration field along the thickness of the sheet induces differential swelling that can produce bending \cite{Holmes2011,Lucantonio2012,AL2013}. However, while differential swelling causes the almost immediate bending of the edges observed experimentally, the central region stays flat until the onset of wrinkling. The analogy between the longitudinal stress $\sigma_x$ and an effective elastic foundation of stiffness $\sigma_x/L_0^2$ \cite{CerdaMahadevan2003} allows us to rationalize the difference in behavior between the edges and the center of the sheet: because the magnitude of $\sigma_x$ is greater in the center of the sheet than at its free edges (Fig.~\ref{fig:contours_stress_preload}b and \ref{fig:p_along_thick}b), the bending of the sheet induced by differential swelling in the central region is initially hampered, while the edges can bend upwards. As solvent absorption proceeds, the longitudinal stress decreases favoring non-flat configurations of the sheet, and the development of transverse compressive stresses (Fig.~\ref{fig:p_along_thick}c) triggers the buckling instability.

\begin{figure}[!t]
\centering
\includegraphics[scale=0.8]{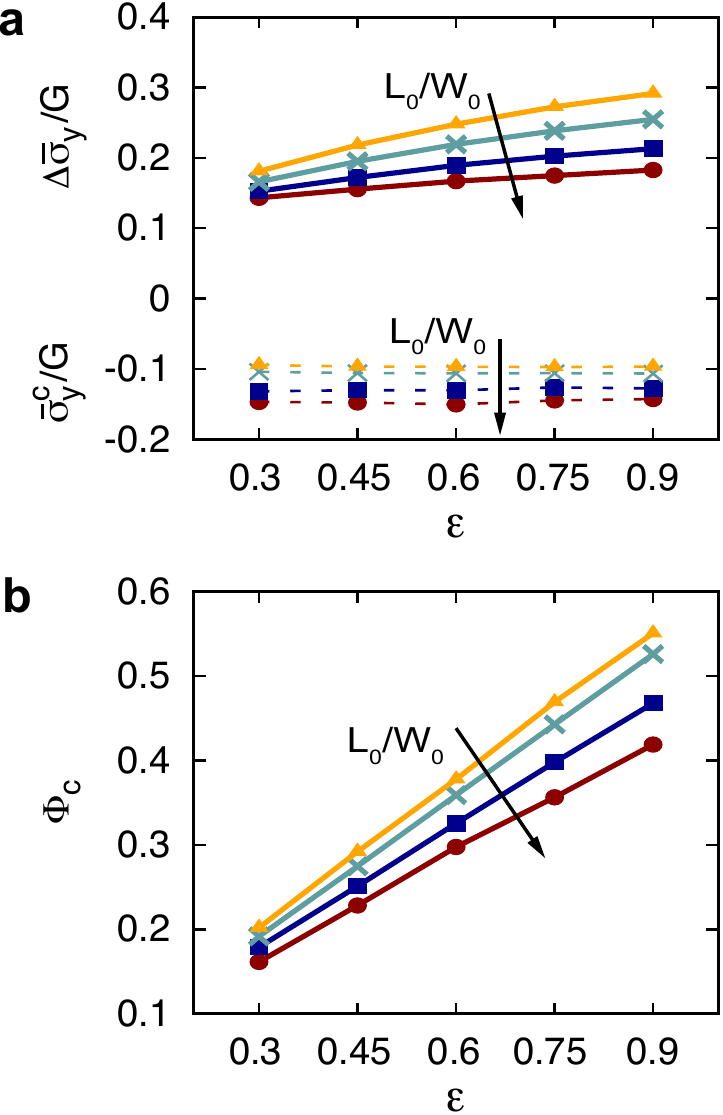}
\caption{Characterization of the wrinkling onset as a function of the applied strain and the aspect ratio $L_0/W_0$ of the unstretched sheet. The data are extracted from numerical simulations. (a) Average transverse stress $\overline{\sigma}_{y}^c$ along the width at wrinkling onset (dashed lines) evaluated at $z/h_0=0$ (bottom surface), and variation of the average transverse stress $\Delta \overline{\sigma}_{y} = \overline{\sigma}_{y}^0 - \overline{\sigma}_{y}^c$ (continuous lines) between the stress  $\overline{\sigma}_{y}^0$  before swelling and the critical stress $\overline{\sigma}_{y}^c$, for different values of the initial aspect ratio $L_0/W_0$: 1 (circles), 5/6 (squares), 5/7 (crosses), 5/8 (triangles). (b) Critical solvent volume fraction $\Phi_c = V_c/V_e$, with $V_c$ the volume of solvent absorbed until the wrinkling onset and  $V_e$ the volume of the dry elastomer.}
\label{fig:onset}
\end{figure}

We obtained physical insights on the sheet state at instability threshold by studying the interplay between solvent migration and rubber elasticity. Solvent diffusion within the polymeric network results in a build-up of an osmotic pressure $p$ along the thickness of the sheet, the pressure being higher in the most swollen layers (Fig.~\ref{fig:p_along_thick}d). 
In turn, the pressure distribution affects both the chemical potential 
\[ \mu(c,p) = \frac{\partial \varphi}{\partial c} + \Omega p \]
of the solvent inside the elastomer, where $\varphi(c)$ is the polymer-solvent free energy of mixing, and the stress within the sheet
\[ \bm{\sigma} = \frac{G}{\lambda_0}\,\frac{\vt{b}}{J}-p\vt{I},\] 
with \mbox{$\vt{b} = \vt{F}\vt{F}^T$} the left Cauchy--Green deformation tensor, \mbox{$\lambda_0 \simeq 1$} is the initial swelling stretch of the sample\footnote{In numerical simulations, a small initial swelling is necessary to avoid the singularity of the mixing free energy \cite{AL2013}.} and 
\[ J=\det \vt{F}=\frac{1}{\lambda_0^3} + \Omega c \]
the swelling ratio. Through this mechanism called stress-diffusion coupling \cite{AL2013}, swelling imposes that the sheet expands laterally against the compressive stretch due to the Poisson effect, and eventually the transverse stress becomes compressive (Fig.~\ref{fig:p_along_thick}c). 

Simple estimates of the axial stretches help to explain better the development of compressive stresses. Let us focus on the central \mbox{($x=L_0/2$)} segment parallel to the transverse direction and located on the bottom surface of the sheet. Before swelling, the longitudinal stretch $\lambda_x \simeq 1+\varepsilon$, while \mbox{$\lambda_z \simeq \lambda_y \simeq 1 - \varepsilon/2 < 1$}; hence, $\lambda_y<\lambda_x\lambda_z$. We expect the axial stretches to increase with swelling; moreover, we verified numerically that the latter relation among stretches still holds during the swelling transient until the onset of wrinkling. Therefore, if we neglect shear deformations, \mbox{$b_y/J \simeq \lambda_y^2/\lambda_x \lambda_y\lambda_z$}: this ratio decreases as the swelling ratio $J\simeq\lambda_x\lambda_y\lambda_z$ increases with solvent absorption, while $p$ increases in the bottom layers, as shown in Fig.~\ref{fig:p_along_thick}d. So, the pressure term in the transverse stress $\sigma_y = (G/\lambda_0)\,b_y/J - p$ becomes dominant and $\sigma_y$ attains negative values all along the width. Wrinkles will appear once the compressive stress reaches a critical average value $\overline{\sigma}_y^c$. Because of the stress distribution in the sheet (Fig.~\ref{fig:p_along_thick}c), the wrinkles emerge first near the edges, as the transverse stress reaches a minimum in these regions. We found that the critical average transverse stress $\overline{\sigma}_{y}^c$ is independent of the applied strain---suggesting that the critical stress is an intrinsic feature of the system,  as usually occurs in buckling problems in structural mechanics---and its modulus increases as the aspect ratio $L_0/W_0$ increases, consistent with the trend observed for the critical stress of transversely loaded plates  (Fig.~\ref{fig:onset}a). Finally, since the initial average transverse stress $\overline{\sigma}_{y}^0$ increases with the applied strain (Fig.~\ref{fig:contours_stress_preload}c), the change in transverse stress $\Delta \overline{\sigma}_{y} = \overline{\sigma}_{y}^0 - \overline{\sigma}_{y}^c$ to reach the wrinkling threshold increases with $\varepsilon$. Hence, because the absorption of solvent during the swelling transient produces such a change in transverse stress (Fig.~\ref{fig:p_along_thick}c), a larger amount of solvent is required to induce a larger change in stress as the applied strain increases (Fig.~\ref{fig:onset}b).

%

\begin{figure}[!t]
\begin{center}
\includegraphics[scale=.9]{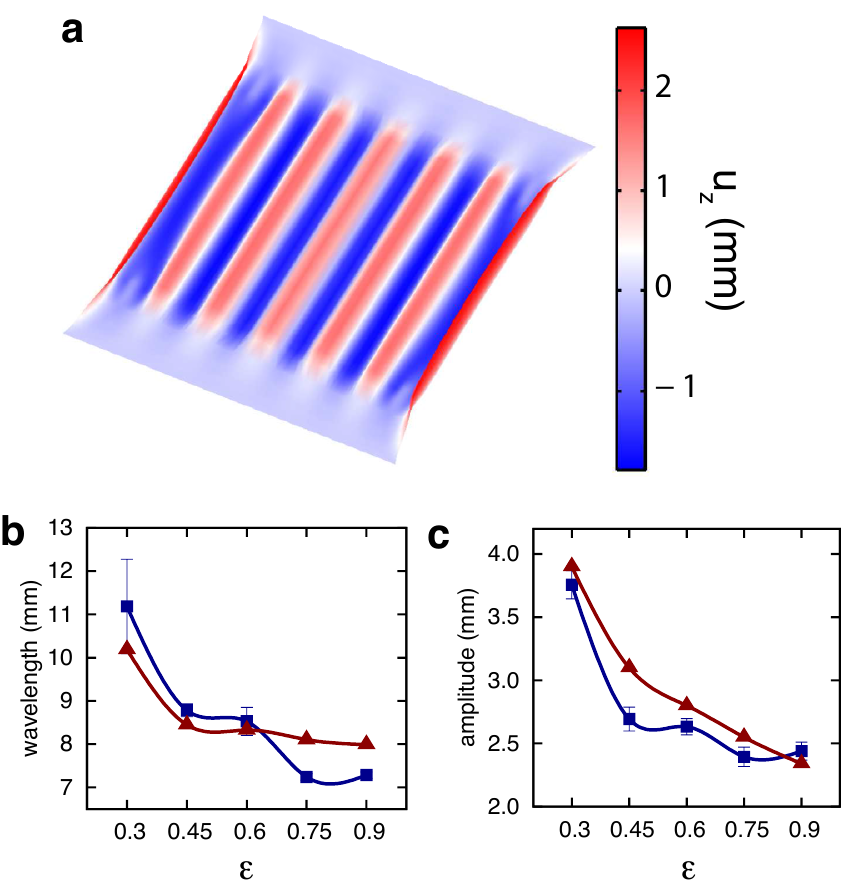}
\caption{(a) Contour plot of the final shape of the swollen stretched elastomer for $\varepsilon = 0.45$. Color code: vertical displacement $u_z$. Comparison between the model and experiments: (b) Wavelength and (c) amplitude at the end of the swelling transient as a function of the initial longitudinal strain $\varepsilon$. For the experimental data, squares represent the average values along the width at the center of the sample, while error bars are the standard deviations of the measurements along the width. Triangles: numerical results. Continuous curves are a guide for the eye. }
\label{fig:wlampvsstrain}
\end{center}
\end{figure}

The shape of the elastomer at the end of the experiment, $t=t_f$, obtained through the numerical model, is depicted in Fig.~\ref{fig:wlampvsstrain}a. The model predicts a decrease of both the wavelength and the amplitude of the wrinkles with an increase of the initial applied strain (\mbox{Fig.~\ref{fig:wlampvsstrain}b--c}), in good quantitative agreement with what we measured experimentally. The discrepancy may be due to a disagreement between the behavior of the edges of the sheets in simulation and in experiments, which affects the growth of the wrinkles. Indeed, in numerical simulations the free edges remain curved upwards (Fig.~\ref{fig:wlampvsstrain}a), while they bend downwards in the experiments as new wrinkles form (see Movie S1$^\dag$). We also found, numerically, that for a sheet with an initial aspect ratio $L_0/W_0 = 5/8$ the trend of the wavelength as a function of the applied strain is qualitatively similar to the one shown in Fig.~\ref{fig:wlampvsstrain}b, but the wavelength slightly increases at each strain (see the ESI$^\dag$).

\section*{Conclusions}

In conclusion, we have presented a study combining experiments and theory in which a uniaxially stretched elastomeric sheet buckles in the direction transverse to the applied stretch when it is exposed to a solvent. The growth of the buckling pattern results from the  swelling of the elastomer by the solvent: the stress field, which was initially tensile everywhere in the stretched sheet, becomes compressive along the transverse direction, and the sheet becomes unstable. The amplitude and the wavelength of the wrinkles depend on the longitudinal strain applied before solvent exposure and on the degree of swelling. For a given nominal strain, the wavelength remains almost constant during the transient swelling process, while the amplitude grows with time. We have shown that a model based on the coupling between solvent diffusion and the state of stress within the sheet captures very well the experimental observations. Our study opens perspectives on the control, through the coupling of strain and swelling, of the dynamic shape of a structure made of a homogeneous polymeric material. 

\section*{Acknowledgments}
A.L.~and P.N.~thank MIUR for support (PRIN 2009 Project No.~200959L72B). H.A.S.~acknowledges the NSF for partial support of this research (CBET-1132835).





\footnotesize{
\bibliography{bibliography} 
\bibliographystyle{rsc} 
}

\end{document}